\documentstyle[a4,11pt]{article}
\begin{document}

\def\thefootnote{\fnsymbol{footnote}}
\def\diam{{\hbox{\hskip-0.02in
\raise-0.090in\hbox{$\displaystyle\bigvee$}\hskip-0.200in
\raise0.099in\hbox{ $\displaystyle{\bigwedge}$}}}}
\def\dop{\mathop{{\diam}}\limits}
\def\bw#1#2#3#4{
{\scriptstyle{#4}}\,
{\dop_{#3}^{#1}}
{\scriptstyle{#2}}  }
\overfullrule=0pt
\def\w#1#2#3#4#5{w \left( {{{#1 \hskip 0.2cm #2} \atop {#3 \hskip 0.2cm #4}} \bigg\vert } {#5} \right)}
\def\wl#1#2#3#4#5{{\overline w} \left( {{{#1 \hskip 0.2cm #2} \atop {#3 \hskip 0.2cm #4}} \bigg\vert } {#5} \right)}
\def\bb#1#2#3#4{B_{\phi_{#1},\phi_{#2}} \left[{{\phi_{1} \phi_{1} }\atop {\phi_{#3} \phi_{#4}}}\right] }

\def\w#1#2#3#4#5{w \left( {{{#1 \hskip 0.2cm #2} \atop {#3 \hskip 0.2cm #4}} \bigg\vert } {#5} \right)}
\def\wl#1#2#3#4#5{{\overline w} \left( {{{#1 \hskip 0.2cm #2} \atop {#3 \hskip 0.2cm #4}} \bigg\vert } {#5} \right)}
\def\B#1#2#3#4#5#6{B_{#1,#2} \left[{{\phi_{#5} \phi_{#6} }\atop {\phi_{#3} \phi_{#4}}}\right] }

\def\el#1#2#3#4{{[ #1 ] [ #2 ] } \over {[ #3 ] [ #4 ]}}

\begin{center}


\vskip 1 cm

{\large \bf Local State Probabilities of Solvable Lattice Models:}
\vskip 0.5cm
{\bf Relatives of $A_{n}^{(1)}$ Family}

\vskip 1 cm

Ernest Baver 

\vskip 1 cm

{\em Department of Particle Physics\\
The Weizmann Institute\\
Rehovot 76100\\
ISRAEL}

\end{center}

\vskip 1 cm

\begin{abstract}
We present the results for the local state probabilities (LSP) of the solvable lattice models, constructed around rational conformal field theory given by WZW model on $SO(3)_{4 R}=SU(2)_{4 R} / Z_{2}$ together with primary field $\phi_{1}$(symmetric tensor of degree 2).  Some conjectures for the LSP for some higher rank relatives of $A_{n}^{(1)}$ face models are also presented. 

\end{abstract}



\section{Introduction}
It was observed in numerous examples \cite{bax,jimbo1,jimbo2} that local state probabilities (LSP) of exactly solvable models in thermodynamic limit are expressed in terms of the characters of the conformal field theory. On the other hand the characters of the conformal field theory play the key role in completely different context \cite{cardy}, namely they enter as the building blocks into modular invariant partition function.  Attempts to achieve the unified understanding of these issues were made \cite{gepner1,thacker, miwa}, but the overall picture is still obscure.

In this paper we present the results for the local state probabilities of exactly solvable lattice models constructed around rational conformal field theory (RCFT) given by WZW model on $SO(3)_{4 R}=SU(2)_{4 R} / Z_{2}$ together with the primary field $\phi_{1}$.

Boltzmann Weights of this model \cite{baver} in the critical limit  were found using the ansatz suggested at \cite{gepner}: 

$$\w{a}{b}{c}{d}{u} \equiv \bw{a}{b}{d}{c}=\sum_{j} \langle a,b,d | P^{(j)} | a,
c,d \rangle \rho_{j} (u), \eqno(1.1)$$

$$\langle a,b,d | P^{(j)} | a,c,d \rangle = \prod_{j \not = l} {{\B{b}{c}{a}{d}{1}{1}-\delta_{b,c} \lambda_{j}} \over {\lambda_{l}-\lambda_{j}} }.\eqno(1.2)$$Here $\B{b}{c}{a}{d}{1}{1}$ is the braiding matrix of the WZW model on $SO(3)_{4 R}=SU(2)_{4 R} / Z_{2}$, $j$ labels the field exchanged in the u-channel and $\rho_{j}(u)$ are some scalar functions depending on the conformal dimensions of the primary fields given by (See \cite{gepner} for the general case):

$$\rho_{0} (u)={{\sin(\lambda-u) \sin(\omega-u)} \over {\sin(\lambda) \sin(\omega)}}, \hskip 0.8cm \rho_{1} (u)={{\sin(\lambda+u) \sin(\omega-u)} \over {\sin(\lambda) \sin(\omega)}}, \eqno(1.3)$$

$$\rho_{2} (u)={{\sin(\lambda+u) \sin(\omega+u)} \over {\sin(\lambda) \sin(\omega)}},   \eqno(1.4) $$where $\lambda,\omega$ are the crossing parameters of the 
model that are related to the conformal weights of the fields appearing in the operator product expansion of field $\phi_{1}$ with itself: $\phi_{1} \times \phi_{1}={\bf 1}+\phi_{1}+\phi_{2}$

$$\lambda={\pi \over 2} (\Delta_{1}-\Delta_{0})={{\pi} \over {4 R+2}}, \hskip 0.8cm \omega={\pi \over 2} (\Delta_{2}-\Delta_{1})=2 \lambda, \eqno(1.5)$$where $\Delta_{j}$ are conformal weights of the primary fields. Projection operators $P^{(j)}$ are found from the braiding matrix $\B{b}{c}{a}{d}{1}{1}$ whose eigenvalues $\lambda_{j}$ are given by $\lambda_{j}=(-1)^{j} e^{i \pi (\Delta_{j}-2 \Delta_{\phi_{1}})}$. Note that the Boltzmann weights vanish, unless the  admissibility condition is satisfied:

$$N_{a,\phi_{1}}^{b} N_{b,\phi_{1}}^{c} N_{c,\phi_{1}}^{d} N_{d,\phi_{1}}^{a} >0, \eqno(1.6) $$where $N_{j,k}^{l}$ are the fusion coefficients. The full elliptic solution of Yang Baxter equation which reduce in the critical limit to the trigonometric solution discussed above was found in \cite{baver} (See Sec.2).

In sequel we will often refer to the related model described in \cite{jimbo2,jimbo3}, which may be considered \cite{gepner} as built around $SU(2)_{k=4 R}$ together with primary field $\phi_{1}$ (symmetric tensor of degree 2). We will call this model diagonal one.

The free energy and LSP of diagonal model were obtained at \cite{jimbo1}. It was shown that the free energies of the lattice model under consideration and diagonal one  are equal \cite{gepner1}. The LSP turns out to be different, but related in a way which will be described below\footnote{I am grateful to Doron Gepner for informing me about his conjectures for the LSP}.

\section{Boltzmann Weights of the Model}

In this section we will recall the definition of the model \cite{baver}. The lattice variables of the model under consideration are in one to one correspondence with the primary fields of the rational conformal field theory given by the WZW model on $SO(3)_{4 R}=SU(2)_{4 R} / Z_{2}$. The lattice variables may take the following values\footnote{The WZW model on $SO(3)_{4 R}=SU(2)_{4 R} / Z_{2}$ has extended current algebra. The fixed point fields $\phi_{R}, \phi_{R^{\prime}}$ have the same conformal weight and isospin, but differ by some additional quantum number}: ${0,1,...,R,R^{\prime}}$

The admissibility condition for the adjacent variables $a,b$ is defined by the fusion coefficients $N_{a,\phi_{1}}^{b}$, where $\phi_{1}$ is the primary field with the lowest conformal dimension. Namely $a \sim b$ iff $N_{a,\phi_{1}}^{b} \not = 0$. The graph corresponding to incidence matrix $N_{a,\phi_{1}}^{b}$ is shown at Fig.1.

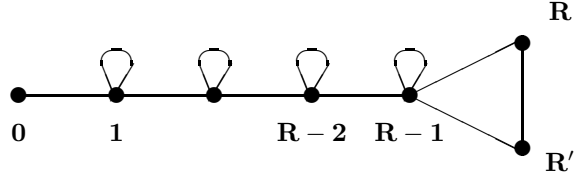
\begin{figure}[ht]
\begin{center}
\unitlength=1.00mm
\linethickness{0.8pt}
\begin{picture}(129.60,92.00)
\put(33.00,80.00){\line(1,0){51.00}}
\put(85.00,80.00){\line(2,1){15.00}}
\put(85.00,80.00){\line(2,-1){15.00}}
\put(100.00,72.00){\line(0,1){16.00}}
\put(85.00,80.00){\circle*{2}}
\put(72.00,80.00){\circle*{2}}
\put(59.00,80.00){\circle*{2}}
\put(46.00,80.00){\circle*{2}}
\put(33.00,80.00){\circle*{2}}
\put(100.00,87.00){\circle*{2}}
\put(100.00,73.00){\circle*{2}}
\put(85.00,84.00){\oval(4,4)[t]}
\put(83.00,84.30){\line(2,-5){2.00}}
\put(84.30,79.30){\line(3,5){3.00}}
\put(72.00,84.00){\oval(4,4)[t]}
\put(70.00,84.30){\line(2,-5){2.00}}
\put(71.30,79.30){\line(3,5){3.00}}
\put(59.00,84.00){\oval(4,4)[t]}
\put(57.00,84.30){\line(2,-5){2.00}}
\put(58.30,79.30){\line(3,5){3.00}}
\put(46.00,84.00){\oval(4,4)[t]}
\put(44.00,84.30){\line(2,-5){2.00}}
\put(45.30,79.30){\line(3,5){3.00}}
\put(105.00,90.00){\makebox(0,0)[cb]{{\small ${\bf R}$}}}
\put(105.00,70.00){\makebox(0,0)[cb]{{\small${\bf R^{\prime}}$}}}
\put(85.00,74.00){\makebox(0,0)[cb]{{\small${\bf R-1}$}}}
\put(72.00,74.00){\makebox(0,0)[cb]{\small ${\bf R-2}$}}
\put(46.00,74.00){\makebox(0,0)[cb]{\small ${\bf 1}$}}
\put(33.00,74.00){\makebox(0,0)[cb]{\small ${\bf 0}$}}

\end{picture}
\caption{Admisibility graph }

\end{center}
\end{figure}

The primary fields (lattice variables) at the Fig.1 correspond to bold points and take values (from the left to the right): $0,1,...,R-1,R,R^{\prime}$, where $R,R^{\prime}$ sit on the base of the right triangle.

 Boltzmann Weights that do not contain fields corresponding to the fixed point were found to be identical \cite{baver} to those of $A_{1}$ model related to symmetric tensor of degree 2 \cite{jimbo1} or $B_{1}$ model related to the vector representation \cite{jimbo4}.

$$\Theta_{1} (u,p)=2 p^{1 \over 4} \sin{u} \prod_{n=1}^{\infty} (1-2 p^{2 n} \cos(2 u) +p^{4 n}) (1-p^{2 n}) \equiv [u], \eqno(2.1)$$

$$\bw{j}{j+1}{j+2}{j+1}={\el{\lambda+u}{\omega+u}{\lambda}{\omega}}, \eqno(2.2)$$

$$\bw{j}{j+1}{j+1}{j+1}={\el{\lambda+u}{(j+1) \omega-u}{\lambda}{(j+1) \omega}}, \eqno(2.3)$$

$$\bw{j}{j}{j+1}{j}={\el{\lambda+u}{j \omega+u}{\lambda}{j \omega}}, \eqno(2.4)$$

$$\bw{j}{j+1}{j+1}{j}={{[\lambda+u][u] \over [\lambda][\omega]}} {{\sqrt{[(j+2)\omega][j \omega]} \over [(j+1) \omega]}}. \eqno(2.5)$$

$$\bw{j}{j+1}{j}{j-1}={\el{u}{\lambda+u -\omega}{\lambda}{\omega}} {{\sqrt{[(j+{3 \over 2}) \omega][(j-{3 \over 2}) \omega]}} \over [(j+{1 \over 2}) \omega]}, \eqno(2.6)$$

{\small $$ \bw{j}{j+1}{j}{j+1}= {\el{\lambda-u}{(2 j+1) \omega-u}{\lambda}{(2 j+1) \omega}}+{\el{u}{(2 j+{3 \over 2}) \omega-u)}{\lambda}{(2 j+1) \omega}} {[j \omega] \over [(j+1) \omega]}, \eqno(2.7)$$}

{\small $$ \bw{j}{j-1}{j}{j-1}= {\el{\lambda+u}{2 j \omega+u}{\lambda}{2 j \omega}}-{\el{u}{2 j \omega+\lambda+u)}{\lambda}{2 j  \omega}} {[(j-{1 \over 2}) \omega] \over [(j+{1 \over 2}) \omega]}, \eqno(2.8)$$}

{\small $$\bw{j}{j}{j}{j}={\el{\lambda-u}{(j+{1 \over 2}) \omega-u}{\lambda}{(j+{1 \over 2}) \omega}}+$$
$$+{\el{u}{(j+1) \omega-u}{\lambda}{(j+{1 \over 2}) \omega}} ({\el{j \omega}{(j+{3 \over 2}) \omega}{(j+1) \omega}{(j+{1 \over 2}) \omega}}+{\el{(j+1) \omega}{(j-{1 \over 2}) \omega}{j \omega }{(j+{1 \over 2}) \omega}}), \eqno(2.9)$$}

$$\lambda={{\pi} \over {4 R+2}}={\omega \over 2}. \eqno(2.10)$$Boltzmann Weights containing fixed point fields and different\footnote{Boltzmann Weights that are not listed here may be obtained from the crossing relation $$\w{a}{b}{c}{d}{u} = \sqrt{{G_{b} G_{d}} \over  {G_{a} G_{c}}} \w{d}{a}{b}{c}{-\lambda-u}, \hskip 0.8cm G_{j} \equiv (2-\delta_{j,R}-\delta_{j,R^{\prime}}) [(j+{1 \over 2}) \omega], $$} from those in \cite{jimbo1,jimbo2} are given by:

{\small $$ \bw{R-1}{R^{\prime}}{R-1}{R^{\prime}}={1 \over 2} {\el{\lambda-u}{(2 R-1) \omega-u}{\lambda}{(2 R-1) \omega}}+{1 \over 2} {\el{u}{(2 R-{1 \over 2}) \omega-u)}{\lambda}{(2 R-1) \omega}} {[(R-1) \omega] \over [R \omega]}+$$
$$+{1 \over 2} {\el{\lambda+u}{\omega+u}{\lambda}{\omega}}, \eqno(2.11)$$

$$\bw{R-1}{R^{\prime}}{R-1}{R}={1 \over 2} {\el{\lambda-u}{(2 R-1) \omega-u}{\lambda}{(2 R-1) \omega}}+$$
$$+{1 \over 2} {\el{u}{(2 R-{1 \over 2}) \omega-u)}{\lambda}{(2 R-1) \omega}} {[(R-1) \omega] \over [R \omega]}-{1 \over 2} {\el{\lambda+u}{\omega+u}{\lambda}{\omega}}, \eqno(2.12)$$}

$$\bw{R}{R-1}{R^{\prime}}{R-1}={\el{\lambda+u}{\omega-u}{\lambda}{\omega}}, \eqno(2.13)$$

{\small $$\bw{R^{\prime}}{R}{R^{\prime}}{R}={\el{\lambda-u}{(R+{1 \over 2}) \omega-u}{\lambda}{(R+{1 \over 2}) \omega}}+$$
$$+{\el{u}{(R+1) \omega-u}{\lambda}{(R+{1 \over 2}) \omega}} ({\el{R \omega}{(R+{3 \over 2}) \omega}{(R+1) \omega}{(R+{1 \over 2}) \omega}}+{\el{(R+1) \omega}{(R-{1 \over 2}) \omega}{R \omega }{(R+{1 \over 2}) \omega}}). \eqno(2.14)$$}There are also other Boltzmann weights containing fixed point, but they may be obtained from Eqs.(2.2-2.10) by replacing in RHS $R,R^{\prime}$ by $R$ in appropriate places, for example

$$\bw{R}{R^{\prime}}{R-1}{R^{\prime}}={\el{\lambda+u}{R \omega+u}{\lambda}{R \omega}}. \eqno(2.15)$$


There are four critical regimes to be considered

$${\rm Regime \hskip 0.3cm I} \hskip 0.8cm  -1<p<0, \hskip 0.5cm 0<u<2 R;$$ 
$${\rm Regime \hskip 0.3cm II} \hskip 0.8cm  0<p<1, \hskip 0.5cm 0<u<2 R;$$ 
$${\rm Regime \hskip 0.3cm III} \hskip 0.8cm   0<p<1, \hskip 0.5cm -1<u<0;$$ 
$${\rm Regime \hskip 0.3cm IV} \hskip 0.8cm  -1<p<0, \hskip 0.5cm -1<u<0.$$In the subsequent sections we use the following notations.

$$E(z,q) \equiv \prod_{k=1}^{\infty} (1-z q^{k-1}) (1-{q^{k} \over z}) (1-q^{k}), \eqno(2.16)$$
$$\Theta_{j,m}^{(\epsilon_{1},\epsilon_{2})}(z,q)=\sum_{n \in Z \atop \gamma=n+{j \over {2 m}}} \epsilon_{2}^n q^{m \gamma^2} (z^{-m \gamma}+\epsilon_{1} z^{m \gamma}), \eqno(2.17)$$

$$\eta (\tau)=q^{1 \over 24} \prod_{k=1}^{\infty} (1-q^{k}), \hskip 0.7cm q=e^{2 \pi i \tau}. \eqno(2.18)$$

\begin{table}
\begin{center}
\begin{tabular}{|l|c|c|c|c|}\hline
Regime & I & II & III & IV \\ \hline
$p$ & $-e^{-{\epsilon \over {4 R+2}}}$ & $e^{-{\epsilon \over {4 R+2}}}$ & $e^{-{\epsilon \over {4 R+2}}}$ & $-e^{-{\epsilon \over {4 R+2}}}$ \\ \hline
$x$ & $e^{-{{2 \pi^2} \over \epsilon }}$ &$e^{-{{4 \pi^2} \over \epsilon }}$  & $e^{-{{4 \pi^2} \over \epsilon }}$ & $e^{-{{2 \pi^2} \over \epsilon }}$ \\ \hline
$\sigma$ & $1$ & $-1$ & $1$ & $-1$ \\ \hline
$q$ & $x^{4 R}$ & $x^{4 R}$ & $x^{2}$ & $x^{2}$ \\ \hline
$u_{a}$ & $E(x^{2 a+1},-x^{2 R+1})$ & $E(x^{2 a+1},x^{4 R+2})$   & $E(x^{2 a+1},x^{4 R+2})$ & $E(x^{2 a+1},-x^{2 R+1})$ \\ \hline
$\lambda_{a}$ & $(2 a+1)(a-2 R)$ & ${1 \over 2}(2 a+1)(2 a-1-4 R)$ & 0 & $a+{1 \over 2}$ \\ \hline
\end{tabular}
\medskip
\caption[table1]{}
\end{center}
\end{table}

\section{Local State Probabilities}
\def\Conf#1#2#3#4{X_{#1}(#2|#3,#4)}           

\subsection{Configuration Sums in Regimes II-III}

 Following arguments of the appendix A of \cite{bax} one may reduce the computation of the local state probabilities to the evaluation of the one dimensional configuration sum $X_{m} (a|b,c;q)$. In particular the local state probabilities are given by

$$P_{m} (a|b,c)={u_{a} x^{\lambda_{a}} X_{m} (a|b,c;q^{\sigma}) \over {\sum_{a} x^{\lambda_{a}} u_{a} X_{m} (a|b,c;q^{\sigma})}}, \eqno(3.1)$$where $u_{a},x,q,\lambda_{a},\sigma$ are defined in the Table 1. In order to find this configuration sum we have to consider the Boltzmann Weights in the one dimensional limit. For our model in regimes II-III we find 

$$\bw{R-1}{R^{\prime}}{R-1}{R^{\prime}}=\bw{R-1}{R}{R-1}{R}={1 \over 2} (1+q), \eqno(3.2)$$ 
$$\bw{R-1}{R}{R-1}{R^{\prime}}=\bw{R-1}{R^{\prime}}{R-1}{R}={1 \over 2} (1-q),\eqno(3.3)$$and otherwise $$\bw{a}{b}{c}{d}=\delta_{b,d} q^{{1 \over 2} |a-c|}. \eqno(3.4)$$One may observe that corner transfer matrix is not diagonal in this limit, but can be easily diagonalized. The configuration sum relevant for the evaluation of the local state probability in this regime is given by

$$\Conf{m}{a}{b}{c}=\sum q^{\sum_{j=1}^{m} j H(l_{j},l_{j+1},l_{j+2}) }, \eqno(3.5)$$where the sum is over all admissible sequences such that $l_{1}=a, l_{m+1}=b, l_{m+2}=c$ and $H(l_{j},l_{j+1},l_{j+2})$    is defined as:

$$H(R-1,R^{\prime},R-1)=1,\hskip 1cm  H(R-1,R,R-1)=0 \eqno(3.6)$$and otherwise

$$H(l_{j},l_{j+1},l_{j+2}) = {1 \over 2} |l_{j}-l_{j+2}|.\eqno(3.7)$$Note that if the fragment of one dimensional sequence contains for example the following pattern $R-1, R^{\prime}, R-1$ or $R-1, R, R-1$, then variables $R,R^{\prime}$ here correspond to the labels of the eigenvalues of the corner transfer matrix, rather then for the spin variables\footnote{The similar situation was encountered for the symmetric tensor $A_{1}$ model in regime IV}. In order to discuss the thermodynamic limit of the configuration sum $\Conf{m}{a}{b}{c}$ we have to specify the ground states $(b,c)$ which are given by:

Regime II: $(b,b+1)$; 

Regime III: $(b,c)$, $b \sim c$ except for $(R^{\prime}, R-1)$ and  $(R, R-1)$

\subsection{Configuration Sums in Regimes I-IV}
We will denote the configuration sum in these regimes by $Y_{m} (a|b,c)$. It is given by the Eq. (3.5) with $H(x,y,z)$ defined by:

{\small $$H(R-1,R^{\prime},R-1)=H(R-1,R,R-1)=H(R,R^{\prime},R)=H(R^{\prime},R,R^{\prime})=0, \eqno(3.8)$$

$$H(R,R-1,R)=H(R^{\prime},R-1,R^{\prime})=2, \eqno(3.9)$$
$$H(R,R-1,R^{\prime})=H(R^{\prime},R-1,R)=1, \eqno(3.10)$$}and otherwise {\small $$H(x,y,z)=H(z,y,x)={\rm Min} (2*(R-y), {1 \over 2} ({\rm Min} (2 x+1,2 z+1)-2 y+1)), \eqno(3.11)$$}where again in the appropriate places in the RHS $R,R^{\prime}$ should be replaced by $R$. Ground states are given by

Regime I: $(R,R^{\prime})$ or $(R^{\prime},R)$;

Regime IV: $(b,c)$, $b \sim c$ and $b+c \le 2 R-2$.

\subsection{Relation with the Diagonal Model}

In order to find the expressions of LSP in terms of modular forms we consider first the expressions, obtained for one dimensional configuration sum of the corresponding diagonal model.

The configuration sums of the corresponding diagonal model in regimes II-III and regimes I-IV are denoted by ${\tilde X}_{m} (a|b,c)$ and  ${\tilde Y}_{m} (a|b,c)$ correspondingly.

 $${\tilde X}_{m} (a|b,c)=\sum q^{\sum {1 \over 2} j |l_{j}-l_{j+2}|}, \eqno(3.12)$$

$${\tilde Y}_{m} (a|b,c) =\sum q^{\sum {j {\tilde H} (l_{j},l_{j+1},l_{j+2})}}, \eqno(3.13)$$
{\small $${\tilde H}(x,y,z)={\tilde H}(z,y,x)={\rm Min} (2*(R-y), {1 \over 2} ({\rm Min} (2 x+1,2 z+1)-2 y+1)), \hskip 0.4cm  y \le R; \eqno(3.14)$$

$${\tilde H}(x,y,z)={\rm Min} (-2*(R-y)-1, {1 \over 2} ({\rm Max} (2 x+1,2 z+1)+2 y+3)), \hskip 0.4cm  y>R  . \eqno(3.15)    $$}In Eqs.(3.12,3.13) the sum is performed over sequences such that {\small $l_{1} \equiv a \sim l_{2} \sim ... \sim l_{m+1} \equiv b \sim l_{m+2} \equiv c$}, and admissibility condition is $a \sim b$ iff $a-b=-1,0,1$ and $a+b=1,2...,4 R-1$.

Let us consider for example configuration sums ${\tilde X}_{m} (a|b,c), X_{m} (a|b,c)$, they are determined completely by the following recursion relations together with the initial conditions\footnote{Note that admissibility conditions in Eqs.(3.16,3.17) are different}:

{\small $${\tilde X}_{m} (a|b,c)=\sum_{d \sim b} {\tilde X}_{m-1} (a|d,b) q^{m {\tilde H} (d,b,c)}, \hskip 0.4cm {\tilde X}_{0} (a|b,c)=\delta_{a,b}, \eqno(3.16)$$}
{\small $$X_{m} (a|b,c)=\sum_{d \sim b} X_{m-1} (a|d,b) q^{m H (d,b,c)}, \hskip 0.4cm X_{0} (a|b,c)=\delta_{a,b}. \eqno(3.17)$$}Using $\tilde X_{m} (a|\sigma(b),\sigma(c))=\tilde X_{m} (\sigma(a)|b,c)$, where $\sigma(a) \equiv 2 R-a$ one may show for $a,b <R$:

{\small $$\tilde X_{m}(\sigma(a)|b,c)=\tilde X_{m} (a|\sigma(b),\sigma(c))=\sum_{d \sim \sigma(b)} \tilde X_{m-1} (a|d, \sigma(b)) q^{m \tilde H(d,\sigma(b),\sigma(c))}= $$

$$=\sum_{\sigma (d^{\prime}) \sim \sigma (b) } \tilde X_{m-1} (a|\sigma (d^{\prime}),\sigma (b)) q^{m \tilde H(\sigma (d^{\prime}),\sigma (b),\sigma (c))}=\sum_{d^{\prime} \sim b } \tilde X_{m-1} (\sigma(a)|d^{\prime},b)) q^{m \tilde H(d^{\prime},b,c)}. \eqno(3.18)  $$}

Summing Eqs.(3.16,3.18) one may observe that $\tilde X_{m}(a|b,c)+\tilde X_{m} (\sigma(a) |b,c)$ obeys the same recursion relation together with initial condition as $X_{m}(a|b,c)$ (See Eq.(3.17)), therefore we conclude\footnote{Note that similar arguments may be applied also in regimes I-IV } that $$X_{m}(a|b,c)=\tilde X_{m}(a|b,c)+\tilde X_{m} (\sigma(a) |b,c).$$ 

Proceeding in the same manner one may arrive to the following relations between the configuration sums\footnote{We list only those relations that are relevant for the evaluation of LSP in thermodynamic limit in specified ground states} $X_{m} (a|b,c), Y_{m} (a|b,c)$ and ${\tilde X}_{m} (a|b,c),{\tilde Y}_{m} (a|b,c)$:

\def\x#1#2#3{X_{m}(#1 |#2,#3)}
\def\xx#1#2#3{{\tilde X}_{m}(#1 |#2,#3)}
\def\y#1#2#3{Y_{m}(#1 |#2,#3)}
\def\yy#1#2#3{{\tilde Y}_{m}(#1 |#2,#3)}

\hskip 5cm Regimes II-III

$$\x{a}{b}{c}=\xx{a}{b}{c}+\xx{2 R-a}{b}{c}, \hskip 1cm a,b,c \not = R, R^{\prime}; \eqno(3.19)$$
$$\x{R}{b}{c}=\x{R^{\prime}}{b}{c}=\xx{R}{b}{c}, \hskip 1cm b,c \not = R, R^{\prime}; \eqno(3.20)$$
{\small $$\x{R}{R-1}{R}=\x{R^{\prime}}{R-1}{R^{\prime}}=\x{R}{R-1}{R^{\prime}}=$$
$$=\x{R^{\prime}}{R-1}{R}=\xx{R}{R-1}{R}, \eqno(3.21)$$ }
{\small $$\x{a}{R-1}{R}=\x{a}{R-1}{R^{\prime}}=$$
$$=\xx{a}{R-1}{R}+\xx{2 R-a}{R-1}{R}, \hskip 0.5cm a \not =R,R^{\prime}; \eqno(3.22)$$}

{\small $$\x{R^{\prime}}{R^{\prime}}{R}=\x{R}{R}{R^{\prime}} \not =\x{R^{\prime}}{R}{R^{\prime}}=\x{R}{{R^{\prime}}}{R} , \eqno(3.23) $$}
$$ \x{R^{\prime}}{R^{\prime}}{R}+\x{R}{R^{\prime}}{R}=\xx{R}{R}{R}, \eqno(3.24)$$ 
$$\x{R^{\prime}}{R^{\prime}}{R}-\x{R}{R^{\prime}}{R}=(-1)^{m}, \eqno(3.25)$$
$$\x{a}{R}{R^{\prime}}=\x{a}{R^{\prime}}{R}=\xx{a}{R}{R}, \hskip 0.5cm a \not =R,R^{\prime}. \eqno(3.26)$$ 
\vskip 1cm
\hskip 5cm Regimes I-IV

$$\y{a}{b}{c}=\yy{a}{b}{c}+\yy{2 R-a}{b}{c}, \hskip 1cm a,b,c \not = R, R^{\prime}; \eqno(3.27)$$
$$\y{a}{R}{R^{\prime}}=\y{a}{R^{\prime}}{R}=\yy{a}{R}{R},\hskip 0.5cm a \not =R,R^{\prime}; \eqno(3.28)$$
$$\y{R}{R^{\prime}}{R}+\y{R}{R}{R^{\prime}}=\yy{R}{R}{R}. \eqno(3.29)$$Note that despite the similarity of the configuration sums $X_{m} (a|b,c), Y_{m} (a|b,c)$ and ${\tilde X}_{m} (a|b,c),{\tilde Y}_{m} (a|b,c)$ the model under consideration and the diagonal one have different admissibility conditions, so that Eqs. (3.19-3.29) represent nontrivial combinatorial identities.

\subsection{Local State Probabilities in Terms of Modular Forms}

Using the expressions found in \cite{jimbo1} for the configuration sums ${\tilde X}_{m} (a|b,c),{\tilde Y}_{m} (a|b,c)$ and Eqs. (3.16-3.26) one may easily obtain the following results:

\vskip 0.7cm
{\bf Regime I:} The system is disordered, namely LSP does not depend on the background configuration. The LSP in the ground state\footnote{By $(R^{\prime},R)+(R,R^{\prime})$ we mean the sum over the LSP that appear in $(R^{\prime},R)$ and $(R,R^{\prime})$ ground states separately} $(R^{\prime},R)+(R,R^{\prime})$ are given by:

$$P_{\rm I} (a)=2 c_{2 a+1} (\tau) T_{2 a+1} (\tau), \hskip 0.8cm a \not =R,R^{\prime} \eqno(3.27)$$
$$P_{\rm I} (R)=P_{\rm I} (R^{\prime})={1 \over 2} c_{2 R+1} (\tau) T_{2 R+1} (\tau), \eqno(3.28)$$ 
$$c_{2 a+1} (\tau)=q^{{(R-a)^2} \over {(2 R+1)}} {E(q^{2 a+1},q^{4 R+2}) \over {\eta (\tau)}}, \eqno(3.29)$$ 
$$T_{2 a+1} (\tau)={{2 \Theta_{2 a+1,2 R+1}^{(-,-)} (x,{x^2}) \eta (\tau)} \over {\Theta_{2 R,2 R}^{(-,-)} (x,{x^2}) \Theta_{0,1}^{(+,+)} (x,{x^2})}}. \eqno(3.30)$$

{\bf Regime II:} Here the LSP in the ground state depends only on $b$

$$P_{\rm II} (a|b)=e_{2 b,2 a}^{4 R} (\tau) T_{2 a+1} (\tau), \hskip 0.8cm a \not =R,R^{\prime}  \eqno(3.31)$$

$$P_{\rm I} (R)=P_{\rm I} (R^{\prime})={1 \over 2} e_{2 b,2 R}^{4 R} (\tau) T_{2 R+1} (\tau), \eqno(3.32)$$

$$T_{2 a+1}(\tau)=x^{{2 R+1} \over 4} {{\Theta_{2 a+1,4 R+2}^{(-,+)} (x,{x^2})} \over \eta(\tau)}. \eqno(3.33)$$

For Regimes III-IV define $s_{1}=b+c, \hskip 0.6cm  s_{2}=b-c+2$
\vskip 0.5cm
{\bf Regime III:} {\small $$ P_{\rm III} (a|b,c)=(c^{(+)}_{s_{1},s_{2},2 a+1} (\tau)+c^{(+)}_{s_{1},s_{2},4 R+1-2 a} (\tau)) T_{s_{1},s_{2},2 a+1} (\tau), \hskip 0.5cm a \not = R,R^{\prime} \eqno(3.34)$$}

$$T_{s_{1},s_{2},2 a+1} (\tau) ={{\Theta_{2 a+1,4 R+1}^{(-,+)} (x,{x^2}) \Theta_{1,2}^{(-,+)} (x,{x^2})} \over {\Theta_{s_{1},4 R}^{(-,+)} (x,{x^2}) \Theta_{s_{2},4}^{(-,+)} (x,{x^2})}} \eqno(3.35)$$

$$P_{\rm III}(R|b,c)=P_{\rm III}(R^{\prime}|b,c)={1 \over 2} (c^{(+)}_{s_{1},s_{2},2 R+1} (\tau) T_{s_{1},s_{2},2 R+1} (\tau), \eqno(3.36)$$

{\small $$P_{\rm III}(R^{\prime}|R^{\prime},R)={{1+c^{(+)}_{s_{1},s_{2},2 R+1}} (\tau) \over 2}  T_{s_{1},s_{2},2 R+1} (\tau), $$
$$ P_{\rm III}(R|R^{\prime},R)={{1-c^{(+)}_{s_{1},s_{2},2 R+1}} (\tau) \over 2}  T_{s_{1},s_{2},2 R+1} (\tau)\eqno(3.37)$$} 

{\bf Regime IV:}

$$P_{\rm IV}(a|b,c)=c^{(-)}_{s_{1},s_{2},2 a+1} (\tau) T_{s_{1},s_{2},2 a+1}, \hskip 0.7cm a \not = R,R^{\prime}, \eqno(3.38)$$

$$P_{\rm IV}(R|b,c)=P_{\rm IV}(R^{\prime}|b,c)={1 \over 2} c^{(-)}_{s_{1},s_{2},2 R+1} (\tau) T_{s_{1},s_{2},2 R+1}, \eqno(3.39)$$ 

$$T_{s_{1},s_{2},2 a+1} (\tau) ={{\Theta_{2 a+1,4 R+1}^{(-,-)} (x,{x^2}) \Theta_{1,2}^{(-,+)} (x,{x^2})} \over {\Theta_{s_{1},2 R-1}^{(-,-)} (x,{x^2}) \Theta_{s_{2},4}^{(-,+)} (x,{x^2})}} \eqno(3.40)$$The branching coefficients $e_{s_{1},s_{2}}^{s_{3}},c_{s_{1},s_{2},s_{3}}^{(+)},c_{s_{1},s_{2},s_{3}}^{(-)}$ are defined below (See for details \cite{jimbo1}).

In Regime II the coefficients $e_{s_{1},s_{2}}^{s_{3}}$ are defined via decomposition of the characters of affine Lie algebra $A_{2 l-1}^{(1)}$ with respect to its subalgebra $C_{l}^{(1)}$:

$$\chi_{\tilde \Lambda_{j}} (q,z_{1},....,z_{l})=\sum_{k=0}^{l} e_{j,k}^{l} (\tau) \chi_{ \Lambda_{k}} (q,z_{1},....,z_{l}). \eqno(3.41)$$In Regimes III-IV the branching coefficients $c_{s_{1},s_{2},s_{3}}^{(+)},c_{s_{1},s_{2},s_{3}}^{(-)}$ are defined via the following identities:

$${{\Theta_{j_{1},m_{1}}^{(-,\epsilon)}(z,q) \Theta_{j_{2},m_{2}}^{(-,+)}(z,q)} \over \Theta_{1,2}^{(-,+)}(z,q)}=\sum_{j_{3}} c_{j_{1},j_{2},j_{3}}^{(\epsilon)} (\tau) \Theta_{j_{3},m_{3}}^{(-,\epsilon)}(z,q), \eqno(3.42)$$where the sum in Eq.(3.42) is over $j_{3}$ such that $0<j_{3}<m_{3}, \hskip 0.5cm   (-1)^{2 j_{3}}=(-1)^{2 j_{1}}$ and $j_{3} \not = m_{3}$ for $\epsilon=+$ and $m_{1},m_{2},m_{3}$ are given by:

$$m_{1}=4 R, m_{2}=4, m_{3}=4 R+2, \hskip 0.6cm ({\rm Regime \hskip 0.3cm III}), \eqno(3.43)$$ 
$$m_{1}=2 R-1, m_{2}=4, m_{3}=2 R+1, \hskip 0.6cm ({\rm Regime \hskip 0.3cm IV}). \eqno(3.44)$$

\subsection{Relation with Conformal Field Theory}

An automorphic property of local state probabilities enables to compute the critical behavior of various order parameters. In regime III the critical exponents of solvable lattice models provide realizations of anomalous dimensions of minimal models \cite{BPZ} of conformal field theory \cite{huse,jimbo5}.   

For example LSP of diagonal model are expressed \cite{jimbo5} in terms of the characters of superconformal minimal models generated by the coset construction {\small $SU(2)_{k} \times SU(2)_{2} / SU(2)_{k}$} \cite{GKO}, giving realization to the model corresponding to the diagonal $(A,A)$ modular invariant \cite{cappelli}.

The LSP of the model considered in this paper in regime III  are given by  (See Eqs. 3.34-3.37):

{\small $$ P_{\rm III} (a|b,c)=(c^{(+)}_{s_{1},s_{2},2 a+1} (\tau)+c^{(+)}_{s_{1},s_{2},4 R+1-2 a} (\tau)) T_{s_{1},s_{2},2 a+1} (\tau), \hskip 0.5cm a \not = R,R^{\prime} $$}Note that the combinations $(c^{(+)}_{s_{1},s_{2},2 a+1} (\tau)+c^{(+)}_{s_{1},s_{2},4 R+1-2 a} (\tau))$ are equal to the characters of the superconformal minimal models labeled by $(A_{4 R-1},D_{2 R+2})$ \cite{cappelli} modular invariant.

\section{Conjectures for the Higher Rank Models}

Consider the $A_{n}^{(1)}$ face models \cite{jimbo3} corresponding to the N-symmetric or N-antisymmetric tensor 
in Regime III: $0<p<1, {-{(n+1)} \over 2}<u<0$. It was conjectured in \cite{jimbo3} that LSP of such models are 
given by

$$P(a)={{b_{\xi,\eta,a}(x^{n+1}) \chi_{a}(x^{n+1},x,...,x)} \over {\chi_{\xi} (x^{n+1},x,...,x) \chi_{\eta} (x^{
n+1},x,...,x)}}, \eqno(4.1)$$where $p=e^{-\epsilon}, x=e^{-{{4 \pi^2} \over L \epsilon}}$ and the identity defining branching coefficients $b_{\xi,\eta,a}$ is:  

$$\chi_{\xi} (q,z_{1},....,z_{n}) \chi_{\eta} (q,z_{1},....,z_{n})=\sum_{a \in L_{l,n}} b_{\xi, \eta, a }(q) \chi_{a} (q,z_{1},....,z_{n}), \eqno(4.2)$$where $\chi_{a}$ is the character of the $A_{n}^{(1)}$ module $L_{a}$ with the highest weight $a$. The equation (4.2) describes the decomposition of the tensor module $L_{\xi} \times L_{\eta}$ ($\xi \in L_{l-M,n}, \eta \in L_{M, n}$) where 
{\small $$L_{l,n}=\{ a=(a_{0},a_{1},....,a_{n}), l+n+1+a_{n}>a_{0}>a_{1}>...>a_{n}, a_{i}-a_{j} \in Z \}. \eqno(
4.3)$$}The identity (4.2) with the choice $M=N$ or $M=1$ is related with  the $A_{n}^{(1)}$ face models corresponding
 to the N-symmetric or N-antisymmetric tensors, respectively.

One may construct models related to the $A_{n}^{(1)}$ face models\footnote{Below we assume for simplicity that $
n+1$ is a prime number} around rational conformal field theory given by WZW model on $SU(n+1)_{k=(n+1) r}/Z_{(n+1)}$ together with the primary field (N-symmetric or N-antisymmetric tensor) starting with the trigonometric ansatz suggested in \cite{gepner} or alternatively using non-critical orbifold procedure suggested in \cite{ginsp}. The admissibility condition now is defined by fusion coefficients of extended current algebra \cite{bavgep} and the lattice variables are restricted to be singlets of $Z_{n+1}$. Denoting the fixed point fields by $R^{j}, j=1,...,n+1$ one may guess the following form of the LSP for these models:

$$P(a)={{\sum_{\sigma \in Z_{n+1}}b_{\xi,\eta,\sigma(a)}(x^{n+1}) \chi_{a}(x^{n+1},x,...,x)} \over {\chi_{\xi} (
x^{n+1},x,...,x) \chi_{\eta} (x^{n+1},x,...,x)}}, \hskip 0.5cm a \not =R^{j}, \eqno(4.4)$$
$$P(R^{j})={1 \over {(n+1)}}{ {b_{\xi,\eta,R}(x^{n+1}) \chi_{a}(x^{n+1},x,...,x)} \over {\chi_{\xi} (x^{n+1},x,
...,x) \chi_{\eta} (x^{n+1},x,...,x)}},  \eqno(4.5)$$where $\sigma$ is the external authomorphism. Note that the combinations $\sum_{\sigma \in Z_{n+1}}b_{\xi,\eta,\sigma(a)}$ are again the characters of the coset models {\small $SU(n+1)_{M} \times SU(n+1)_{L-M} / SU(n+1)_{L} $} corresponding to the D modular invariant.

\vskip 0.7cm
\hskip 4.5cm {\bf ACKNOWLEDGMENTS}
\vskip 1cm
I am grateful to Doron Gepner for numerous useful discussions and for informing me about his conjectures for the local state probabilities.


\end{document}